\documentstyle [12pt] {article}
\newcommand{\be}{\begin{eqnarray}}
\newcommand{\ee}{\end{eqnarray}}
\newcommand{\bfl}{\begin{flushleft}}
\newcommand{\efl}{\end{flushleft}}

\newcommand\half{\frac 1 2 }

\begin{document}

\begin{flushleft}
  \hfill USITP-92-4\\
                    \hfill  May  1992 \\
\end{flushleft}
\vspace*{3mm}
\begin{center}

{\Huge\bf  Space-Time Symmetries of\\ \vskip 3mm
Quantized Tensionless Strings}
           {\Large\footnote{Work supported by The Swedish Natural
                               Science Research Council.}} \\
\vspace*{22mm}

 {\Large J. Isberg, U. Lindstr\"{o}m and  B. Sundborg\\ }

\vspace*{11mm}

   Institute of Theoretical Physics \\ University of Stockholm \\
    Vanadisv\"{a}gen 9 \\
    S-11346 Stockholm \\  SWEDEN  \\

\end{center}
\vspace*{25mm}

{\small {\bf Abstract:} The tensionless limit of the free bosonic string is
space-time conformally symmetric classically. Requiring invariance of the
quantum theory in the light cone gauge tests the reparametrization symmetry
needed  to fix this gauge. The full conformal symmetry gives stronger
constraints than the Poincar\'e subalgebra. We find that the symmetry may
be preserved in any space-time dimension, but only if the spectrum is
drastically reduced (part of this reduction is natural in a zero tension
limit of the ordinary string spectrum). The quantum states are required to be
symmetric ({\it i.e.} singlets) under space-time diffeomorphisms, except
for the centre of mass wave function.}

 \vfill

\eject

\setlength{\baselineskip}{20pt}
\begin{flushleft}
\section{Introduction}
\end{flushleft}

\bigskip

The high energy limit of string theory is still quite poorly understood,
despite many
important and interesting results on high energy scattering,
\cite{AMAT3}-\cite{GROS4} and
high temperature behaviour \cite{ALVA1}-\cite{ATIC1}.
Just like the massless limit in particle theory
sheds light on short distance field
theory, the zero tension limit, $T\to 0$,
of strings is expected to illuminate some
short-distance properties of string theory.
In particular we hope that the intriguing high
energy symmetries discussed by Gross \cite{GROS1}
may be studied in this limit. Thereby one
would presumably be able to probe the conjectured
unbroken "topological" phase of
general covariance
\cite{WITT1,WITT2}.
Though much work has been done in topological field theory, \cite
{WITT3,LABA1,WITT4,BIRM1}, much less is known
about the relation to string theory. The
present work supports and substantiates such a connection.
In fact, our results point in this direction in quite an unexpected way. We
have
approached the problem by first formulating
the exactly tensionless case, where additional
symmetries, in particular space-time conformal
symmetry should become manifest. In a
previous series of papers we have studied
various aspects of the tensionless case of the
bosonic string \cite{akul}, the superstring
\cite{ulbsgt1} and the spinning string
\cite{ulbsgt2,roli}. Other authors have also
discussed tensionless strings and their
quantization \cite{zh,banerual,liraspsr,gararual}
ever since they were first discussed by
Schild \cite{sc}.

Some basic questions about the spectrum of tensionless string and its relation
to the spectrum of the ordinary string have not been satisfactorily answered.
For instance, one expects either a continuous
or a massless spectrum when the scale given
by the string tension is removed from the theory. Correspondence with classical
tensionless strings would favour a continuous
spectrum, but on the other hand all $T\not=
0$ string states approach zero mass as $T\to 0$.
 Previous authors have advocated both a
continuous mass spectrum, \cite{LIZZ1,gararual},
and a massless spectrum, \cite{gararual}. We do
not agree with any of the presented
spectra. The result of the
present work indicates that the massless
spectrum is the correct answer, but we also find
extremely restrictive constraints on the
spectrum, effectively allowing only states
invariant with respect to general coordinate
transformations. One could envisage a
spectrum of string states characterized
by topology, but we have not yet found a concrete
construction of a satisfactory Hilbert space.

In this paper we will be concerned with the
quantization of the tensionless closed bosonic
string. In particular we will explore
the (space-time!) conformal symmetry of this model and investigate under what
condition this symmetry survives at the quantum level. When the two-dimensional
reparametrization invariance of ordinary
strings is gauge-fixed in the light-cone
gauge, anomalies in the local symmetry are
reflected in a breakdown of the Lorentz
algebra (in non-critical dimensions).
Similarly, inconsistencies in the quantum
geometry of the tensionless string can be
probed by checking the conformal algebra in the
light-cone gauge. We are further motivated
to demand space-time conformal invariance in
the quantized theory because we find that this is the symmetry that replaces
Weyl-invariance in the $T \to 0$ limit.
Hence when we find obstructions to conformal
invariance, rather than conclude that the symmetry is broken, we interpret the
obstructions as conditions on the physical
states of fundamental strings. In a future
tensionless limit of $QCD$ strings, it might
instead be more appropriate to accept
breakdown of conformal invariance. The
description of our calculations will be very brief,
but we plan to give more details in a later publication.

\bigskip

\begin{flushleft}
\section{The Classical Theory}
\end{flushleft}

\bigskip

There are many formulations of the bosonic null-string. The one that we have
found most useful, and which is the closest analogue of the Weyl-invariant
formulation of the tensionful string, involves a contravariant world sheet
vector density $V^\alpha$ of weight ${\textstyle{1 \over 2}}$, \cite{ulbsgt1}.
Geometrically it describes the degenerate,
($\det g^{\alpha\beta}=0$), limit of the
auxiliary metric density. These formulations
of the tensionful and the tensionless
string  can be derived systematically
from a Hamiltonian formulation of the Nambu-Goto
string. The bosonic zero tension action is
\be
S_1^0=\half\int {d^2\xi V^\alpha} V^\beta\partial _\alpha X^m \partial
_\beta X^n \eta _{mn}\;\; ,\label{a}
\ee
where $X^m (\xi )$ are the space time coordinates for the string, $\xi ^a=(\tau
,\sigma )$ parametrize the world-sheet and
\be
\gamma _{\alpha\beta}=\partial _\alpha X^m \partial
_\beta X^n \eta _{mn}\label{ind.metr}
\ee
is the induced metric on
the world-sheet. The action (\ref{a}) is invariant under two different
symmetries: World-sheet diffeomorphisms and space-time conformal
transformations. Under the diffeomorphisms $X^m$ transforms as a scalar field
and $V^\alpha$ as a vector density:
\be
\delta _\varepsilon V^\alpha =-V \cdot \partial \varepsilon ^\alpha
+\varepsilon \cdot \partial V^\alpha +{\textstyle{1 \over 2}}(\partial \cdot
\varepsilon )V^\alpha .\label{b}
\ee

The field equations that follow from the
action (\ref{a}) are:
\be
V^\beta \gamma _{\alpha \beta}=0,\qquad \partial _\alpha (V^\alpha
V^\beta \partial  _\beta X^m )=0\label{e}
\ee
The first of these equations states that $\gamma_{\alpha\beta}$ has an
eigenvector with eigenvalue zero which implies that it is
a degenerate matrix:
\be
\det \gamma _{\alpha \beta}=0\label{f}
\ee
This means that the world sheet spanned by the tensionless
string is a {\it null surface}. The second of the field
equations is most easily interpreted by
gauge fixing the reparametrization symmetry
(\ref{b}) to the {\it transverse gauge},
\be
V^\alpha = (v,0),\label{g}
\ee
with $v$ a
constant. The equations of motion are then
\be
\ddot X^m &=& 0 \label{xpp}\\
\dot X^2 &=& \dot X \cdot X'=0.\label{eq.mo}
\ee
Clearly the "null"
string behaves classically as a collection of massless
particles, one at each $\sigma$ position, constrained to move
transversally to the direction of the string.

The enlargement of the Poincar\'e
symmetry to conformal symmetry when $T \to 0$
is the same phenomenon that occurs for the
massless particle. In fact, since a conformal
transformation will scale the D-dimensional line element,
it will also scale the induced metric. This can be
compensated by a ($X^m$-dependent) scaling of $V^\alpha$, and the
action (\ref{a}) thus be left invariant. Note that this is not
possible for the action of the tensionful string,
\be
S_1=-{T \over 2}\int {d^2}\xi \sqrt {-\det
g_{\gamma\delta}}g^{\alpha\beta}\gamma _{\alpha\beta},\label{d}
\ee
since any rescaling of
$g_{\alpha\beta}$ is an invariance of $\sqrt {-det
g_{\gamma\delta}}g^{\alpha\beta}$ alone. In fact, in this sense {\it
Weyl-invariance is replaced by conformal  invariance} in the limit $T \to 0$.

Conformal symmetry extends the Poincar\'e symmetry for massless particles and
massless free fields. Similarly
classical tensionless strings should enjoy conformal
symmetry. The infinitesimal transformations form the conformal algebra:
\be
\lbrack M_{mn},M^{rs}] &=& \delta_m^r M_n^{~s} -
\delta_n^r M_m^{~s} + \delta_n^s
M_m^{~r} - \delta_m^s M_n^{~r} \cr
\lbrack M_{mn},P^s] &=& \delta_n^s P_m - \delta_m^s P_n, \qquad
\lbrack P_m,P_n] = 0 \cr
\lbrack M^{mn},K_l] &=& \delta^n_l K^m - \delta^m_l K^n, \qquad
\lbrack K^m,K^n] = 0\cr
\lbrack P_m,K^n] &=& \delta_m^n D -M_m^{~n}\label{c.alg}\\
\lbrack D, M_{mn}] &=& 0\cr
\lbrack D,P_m] &=& - P_m\cr
\lbrack D,K^m] &=&  K^m \nonumber
\ee
Note that the whole algebra can be generated from repeated commutators of
translations $P_m$ and conformal boosts $K^n$.
Therefore it is enough to determine
the transformations $K^n$ discussed below. (The translations are given by the
momenta, act just as expected and do not cause any complications.)

Under Poincar\'e transformations $X^m$ behaves as a
Lorentz vector and $V^\alpha$ as a scalar. In contrast, conformal boosts
(generator $K^m$) and dilations (generator $D$)
rescale $V^\alpha$ in order to leave
the action invariant when the induced metric (\ref{ind.metr}) is rescaled by
the
ordinary action of conformal transformations on the coordinates $X^m$. The full
transformations act as follows:
\be \delta _b X^m = [b\cdot K,X^n]=(b \cdot
X)X^m-\half X^2b^m, \qquad  \delta _b V^\alpha =-b \cdot XV^\alpha\\ \delta _s
X^m=[sD,X^m]=sX^m, \qquad \delta _s V^\alpha= -sV^\alpha \label{c}
\ee

Since $V^\alpha$ is rescaled, the conformal transformations
destroy the transverse gauge unless we modify them by
including a compensating diffeomorphism transformation.
Preservation of the gauge choice requires that the
parameters of this compensating transformation satisfy
\be
\dot \varepsilon ^1=0, \qquad \dot \varepsilon ^0-
 \varepsilon '^1 = - 2 b\cdot X,
\qquad \varepsilon ^a(\tau,\sigma + 1) = \varepsilon
^a(\tau,\sigma)\label{i}
\ee
The last condition arises because we are dealing with the closed string,
parametrized to have period $1$ in $\sigma$.

The transverse gauge (\ref{g}) is not a complete gauge fixing. The gauge choice
is preserved by the following special 2D diffeomorphisms:
\be
\delta \xi ^\alpha = \lambda ^\alpha, \qquad \lambda ^\alpha = (f'(\sigma
)\tau + g(\sigma ),f(\sigma )) \label{h}
\ee
with $f$ and $g$ arbitrary functions of $\sigma $ only. The residual symmetry
(\ref{h}) corresponds to the 2D conformal symmetry of the tensionful string in
conformal gauge.

We introduce light cone coordinates $(X^+ ,X^- ,X^i)$, where $X^\pm \equiv {1
\over {\sqrt 2}}(X^0\pm X^{D-1})$ and $i=1...D-2$. This is
only a choice of coordinates, but next we use the residual
symmetry (\ref{h}), and the equation of motion  (\ref{xpp}),
 to fix a light cone gauge. From
\be
\tilde \tau =f'(\sigma )\tau +g(\sigma), \qquad \ddot X ^+=0
\label{j}
\ee
we see that we may take $\tilde \tau \propto X^+$, and we
choose
\be
X^+=\frac{\pi^+}{v^2}\tau \label{k}
\ee
where $\pi^+$ is a constant. This completely fixes the
diffeomorphism gauge, except for rigid $\sigma$-translations.

In light cone  coordinates the
$V^\alpha$ equations of motion read
\be
V^\alpha \partial _\alpha X^i \partial _\beta X^i
-V^\alpha \partial _\alpha X^- \partial _\beta X^+
-V^\alpha \partial _\alpha X^+ \partial _\beta X^- = 0 ,\label{m}
\ee
and in transverse gauge (\ref{g}) they give the constraints
\be
\dot X^i \dot X^i -2\dot X^- \dot X^+ &=& 0\label{n}\\
 \dot X^i X'^i -\dot X^- X'^+-\dot X^+ X'^- &=& 0\label{n2} .
\ee

We now use (\ref{n},\ref{n2}) in the light cone gauge (\ref{k}) to eliminate
$X^-$, except for a zero-mode $\bar{X}^-(\tau)$:
\be
X'^- &=& {v^2 \over \pi ^+}\dot X^i X'^i\label{gnutt}\\
\dot X^- &=& {v^2 \over 2\pi ^+}\dot X^i \dot X^i \label{fnutt}\\
\bar{X}^- &\equiv& \int d\sigma X^- \equiv
x^- + \frac{1}{v^2}\pi^-\tau \; ,\label{o}
\ee
where  $x^-$ and $\pi^-$ are constants.
Having eliminated $X^\pm $ we are left with the equations of motion
for the transverse components,
$X^i$, (in transverse gauge)
\be
 \ddot X^i=0 \label{p}
\ee
These may be derived from the light cone action
\be
S_{LC}=\frac{v^2}{2}\int {d^2\xi \dot X^i \dot X^i} \label{q}
\ee

Again, conformal transformations will
destroy our gauge choice and we need additional compensating
transformations. They may be found following
the procedure used by Goddard et al.
\cite{ggrt} for the Lorentz transformations
of the ordinary string. The conformal
boosts which preserve the light cone gauge are: \be
\delta^{l.c.}_{b}X^m (\tau,\sigma) & = &(b \cdot X)X^m-\half X^2b^m
    +\epsilon^0 \dot{X}^m +\epsilon^1 X^{'m} \label{l}
\ee
\be
\frac{1}{v^2}\epsilon^0 &=&\frac{1}{2\pi^+}
b^+ X^2_0 + \tau\frac{b^+}{\pi^+} \left(
     X_0\cdot\dot{X}_0  - \int d\sigma' X_0\cdot\dot{X}_0 \right) -\cr
     &-&\tau b\cdot\left( X_0 + \int d\sigma' X_0 \right)
     -\tau^2 b\cdot \dot{X}_0 \label{eps0} \\
\frac{1}{v^2}\epsilon^1 &=& b\cdot\int d\sigma' X_0(\sigma')
h(\sigma' - \sigma) +
             \frac{b^+}{\pi^+} \int d\sigma'
X_0(\sigma')\cdot\dot{X}_0(\sigma')
              h(\sigma' - \sigma)\; ,\label{eps1}
\ee
where
\be
h(\sigma - \tilde \sigma) \equiv
 \sigma -\tilde \sigma -{\textstyle{1 \over
2}{\rm sign} (\sigma -\tilde \sigma ) } \label{ad}
\ee
and $X_0,\dot{X}_0$ are values
taken at $\tau=0$.
Equations (\ref{eps0},\ref{eps1}) satisfy (\ref{i}).

We now wish to find a canonical formulation
of the generators in order to prepare
for quantization.
The transverse conjugate momenta can be read off from (\ref{q}), they
are
\be
\Pi ^i=v^2 \dot X^i \label{r}
\ee
In addition $-\pi^+$ is canonically conjugate to $x^-$.
We now proceed to a Hamiltonian formulation of the theory, and start
by rewriting (\ref{gnutt},\ref{fnutt}) using (\ref{r})
\be
X'^-={1 \over \pi ^+ } X'^i \Pi^i \label{mx}
\ee
\be
\dot X^-={1 \over 2\pi ^+ v^2}\Pi ^i \Pi ^i \equiv \frac{1}{v^2}\Pi^- \label{s}
\ee
Equations (\ref{k},\ref{r}) and (\ref{s}) give the translation operators:
\be
P^m = \int d\sigma \Pi^m \equiv \pi^m \label{gus}.
\ee
The action (\ref{q}) then corresponds to a Hamiltonian
\be
{\cal H}={\textstyle{1 \over {2v^2}}}\int{d\sigma  \Pi^i
\Pi^i}={\textstyle{1 \over {v^2}}}\int{d\sigma \pi ^+ \Pi ^-},
\ee
which indeed generates $\tau$-translations, c.f. (\ref{k},\ref{gus}).

The generators of the conformal group additional to the Lorentz and translation
generators, can now be written (at $\tau =0$):
\be
\!\!\!D &=& \int {d\sigma \left\{ {X^i(\sigma )\Pi ^i(\sigma
)-X^-(\sigma )\pi ^+ } \right\}}\label{u}\\
\!\!\!K^i &=& \int {d\sigma \left\{X^i(\sigma )X^j(\sigma )\Pi ^j(\sigma )
-X^i(\sigma )X^-(\sigma )\pi ^+
-\half X^jX^j(\sigma )\Pi ^i(\sigma ) \right\}}\label{v}\\
\!\!\!K^- &=&\! \int{ \!d\sigma \left\{ X^-(\sigma
)X^j(\sigma )\Pi ^j(\sigma )\!-\!\half X^-(\sigma )X^-(\sigma )\pi ^+
\!-\!\half
X^i(\sigma )X^i(\sigma )\Pi ^-(\sigma ) \right\}}\label{w}\\
\!\!\!K^+ &=&-\half \int{d\sigma \left\{X^i(\sigma )X^i(\sigma )\pi
^+\right\}}\label{x}
\ee
We have checked that these generators indeed
give the transformations in (\ref{l}).

\begin{flushleft}
\section{Quantization}
\end{flushleft}

\bigskip

We quantize the theory by introducing the fundamental commutation relations
\be
\left[ {X^i(\sigma ),\;\Pi ^j(\tilde \sigma )} \right]=i\delta ^{ij}\delta
(\sigma -\tilde \sigma ) \qquad \left[ {X^-,\;\pi ^+} \right]=-i\label{t}
\ee
All other commutation relations between the transverse variables
vanish and those for the composite operator $X^-$  will be derived
below. It is
sufficient to check the conformal algebra
at $\tau =0$, since the algebra itself contains
generators of $\tau$-translations. Hence
we may consider the variables in (\ref{t}) as
functions of $\sigma$ only  (as in (\ref{u}-\ref{x})).
In the tensionful theory one would then expand the operators in
oscillators. The motivation for using this expansion is that each oscillator
mode solves the equations of motion. In the present, tensionless case the
solutions to the equations of motion
(\ref{eq.mo}) are instead given by the local
position and momentum, since each point of
the string moves like a free particle.

Therefore we treat the generators
directly in the functional form (\ref{u}-\ref{x}).
Note that they are highly non-linear,
up to quartic in the fields. The corresponding
ordering problems will lead to lower
order terms and possibly to anomalies of the conformal
algebra. Such anomalies can only show up in commutators of the cubic
generators $K^i,M^{i-}$ and the quartic generator $K^-$.
(The Lorentz generator $M^{i-}$
is the culprit in the tensionful case where it is responsible for anomaly of
the
Lorentz algebra outside the critical dimension).

Commutators of cubic and quartic generators in general produce products of
Dirac
delta functions and other distributions.
To make sense of the quantum symmetry one
therefore has to regularize products of operators at coinciding points. A
simple
way of doing this is to smear the $\sigma$-dependence of $X(\sigma)$ and
$\Pi(\sigma)$ by convoluting them with
some test function approaching a Dirac delta
function. Effectively the delta function
in (\ref{t}) is replaced by a regularized
delta function $\delta _\varepsilon (\sigma)$, which should satisfy:
\be
\mathop{\lim }\limits_{\varepsilon \to 0}
\int {d\sigma f(\sigma )\delta _\varepsilon
(\sigma )}=&f(0) \qquad
\delta _\varepsilon (-\sigma) &= \delta _\varepsilon (\sigma)\\
\int d\sigma \delta _\varepsilon (\sigma) =& 1 \qquad \quad
 \delta _{s\varepsilon} (\sigma) &=
 {1 \over s} \delta _\varepsilon ({\sigma \over s})
\label{af}
\ee
The last assumption facilitates power-counting of divergences in the limit
$\varepsilon \to 0$. It can be used to
exclude a priori possible extra terms in the
generators. The regularization clearly
introduces a world-sheet length scale, which may
cause problems, but the purpose of the
investigation is precisely this: To see whether the
{\it space-time} physics of the model still
makes sense, and in particular whether
conformal symmetry can be preserved.

The check of the conformal algebra in the quantum theory starts
with a derivation of the commutators for the composite operators. We find
(to order $\varepsilon^2$, which is enough for the conformal algebra)
\be
\left[X^-,{1 \over {\pi ^+}}\right]\!\!\! &=&
\!\!\!  {i \over (\pi^+)^2}\label{aa}\\
\left[ {X^i(\sigma ),X^-(\tilde \sigma)} \right]\!\!\!  &=& \!\!\!
{i \over {\pi ^+}}X'^i(\sigma) h(\sigma - \tilde \sigma) \label{z}\\
\left[\Pi^i(\sigma),X^-(\tilde \sigma) \right]
\!\!\!  &=&\!\!\! {i \over {\pi ^+}}
 \left \{ \Pi'^i(\sigma)h(\sigma - \tilde \sigma ) + \Pi^i(\sigma)
(1- \delta_\varepsilon (\sigma -\tilde \sigma ) )  \right\} \label{ab}\\
\left[X^-(\sigma),X^-(\tilde \sigma) \right]\!\!\!  &=&\!\!\!  {i
\over {\pi ^+}}\left\{\left(X'^-(\sigma)+\beta \right)h(\sigma - \tilde \sigma)
\!-\!\left(X'^-(\tilde \sigma)+\beta \right)h(\tilde \sigma -
\sigma)\right\}\label{ac2}
\ee
where
and $\pi^+ \beta$ is the generator of rigid $\sigma$ translations whose
action on an arbitrary function $f(\sigma)$ is
\be
\left[\pi^+ \beta,f\right]=-i f' \\
\beta = {1 \over {\pi^+} } \int d\sigma X'^i  \Pi^i \label{ae}
\ee
(Similar relations were presented in \cite{banerual},
but they do not properly take the
periodicity in $\sigma$ into account.) Clearly the composite operators and
their
commutation relation have to be periodic, since they are derived from periodic
canonical variables. However, the commutation
rules of $X^-$ are only given in terms
of those of $X'^-$, which have to be integrated. The integration constant can
be
fixed from (\ref{aa}), but manifest periodicity of the commutation relations
can
only be achieved by identifying
\be
X'^-={1 \over \pi ^+ }\Pi^i X'^i - \beta ,
\ee
instead of the naive equation (\ref{mx}).
The last term vanishes in the physical phase
space, since rigid $\sigma$-translations are
the last surviving remnant of reparametrization invariance, which is not
fixed in the light-cone gauge. Still one has to keep track of it and in
the quantum theory it cannot be set to zero
until it has been commuted to the left
or right and acts on physical states.

In checking the conformal algebra at the
quantum level we have to ensure {\it hermiticity} of the operators. This we can
do by introducing an additional term $cX^i$ in the expression for $K^i$. The
constant $c$ may be chosen to give a hermitean $K^i$.
In general, $c$ will depend on the space
time dimension, but it turns out that its value
will not affect the algebra.
Then starting from the
hermitean operators $P^-$ and $K^i$ we
"bootstrap" our way to the full algebra by
successively evaluating the commutators
(\ref{c.alg}). In this way all the generator
expressions we derive will automatically be hermitean.

In evaluating the commutators between the generators of the conformal
algebra we choose a reference ordering of $X$ and $\Pi$.
We use an ordering where $X$:s stand to the left of $\Pi$:s, and the
combination
$\pi^+ X^-$, which is convenient to use, is placed in between $X$:s and
$\Pi$:s.
If there are any $\pi^+$:s left they are placed leftmost in a term.
Technically the evaluation proceeds by commuting the operators to this
reference
order, identifying cancellations, discarding terms that are total derivatives
in
$\sigma$ and collecting the remaining terms.
As a warm up, before doing the string calculation we
checked the conformal algebra in the
simpler case of the quantized massless particle.
To our knowledge this calculation has not
been done before. The calculation is simpler due to the lack of
$\sigma$ dependence, but the non-linearities are
there, and a priori there seems to be
ample opportunity for obstructions to arise.
The quantized system turned out to be
conformally invariant without any
restrictions on the physical state-space, however.

Returning to the null string, we started from $P^-$
and $K^i$, calculated $M^{j-}$ and then
$K^-$ using (\ref{c.alg}). Having thus determined
all generators one can continue to check the
algebra. A calculation of the Lorentz
subalgebra verifies the result of \cite{LIZZ1},
that it is satisfied in arbitrary
dimensions without any restrictions. When we
widen the scope to the full conformal
algebra, already $[K^i,M^{j-}]$ causes problems.
This commutator should be proportional to
$\delta^{ij} K^-$ but we find a term  \be
\propto \int{d\sigma}{1 \over {\varepsilon \pi^+}} [X^i \Pi^j - x^i \pi^j]
\equiv {1 \over {\varepsilon \pi^+}}L^{ij}\label{constraint}
\ee
with a different index structure. Apparently, the
conformal algebra is anomalous, unless
one imposes that the Hilbert space of physical
states is constrained by
\be
\tilde L^{ij}|phys> = 0 ,\label{phys}
\ee
where $\tilde L^{ij}$ is the traceless
part of $L^{ij}$, and the trace part is absorbed
into $K^-$. (Note that the zero mode
contribution is subtracted from $X^i \Pi^j$ in
(\ref{constraint}). This explains why no
such restriction is found for the massless
particle; there the zero modes are the whole story). The calculation leading to
(\ref{constraint}) has been done both
using Mathematica and by hand. Clearly one has to
commute the constraint with all conformal
generators to regain a symmetric Hilbert space.
In addition, one has to do the rest of
the conformal algebra. In the process further
constraints will be generated, which should
in their turn be commuted with conformal
generators and other constraints. Eventually,
all constraints close to form some algebra.
Thanks to our Mathematica programme we can
do sufficiently many steps to find this
structure.

Commuting $P^{-}$ with $\tilde L^{ij}$ we
get a new constraint
\be
\int{d\sigma} [\Pi^i \Pi^j -\pi^i \pi^j] = 0 .
\ee
The trace part can be obtained by instead
commuting with $K^-$, and this constraint means
that surviving states are massless. Classically,
the trace is an integral of a square,
implying that the momentum density should
equal its average, i.e. the zero mode. Due to the
uncertainty relation, there should then only
be one corresponding quantum state, which is
completely delocalized, in the non-zero
mode Hilbert space. However, the above argument
hinges on local momentum densities and
does not carry over to regularized expressions.
Though we do not have an explicit construction
of the Hilbert space we believe that
non-trivial solutions should exist.

The constraints $\tilde L^{ij}$
are simply generators of special linear
transformations on non-zero modes in the
transverse directions. By construction
the full set of constraints will be closed under
conformal transformations, and Lorentz transformations in particular. The
tracelessness condition on $\tilde L^{ij}$
is actually covariant since only transverse
dimensions contribute to non-zero modes at $\tau = 0$. Thus the constraints on
non-zero modes are generated by successive
commutators of conformal transformations and
all special linear transformations. We
may now apply a result by Ogievetsky \cite{og},
stating that the set of these transformations
close to the algebra of general coordinate
transformations. States should then correspond to equivalence classes of string
configurations invariant under general
coordinate transformations. For example, various
self-intersecting strings could correspond to different states.

\bfl
\section{Discussion}
\efl

To impose anomalous terms in the algebra as
constraints on the Hilbert space is certainly
unconventional, but it seems to be the only
way to construct a consistent tensionless
string. We can get an idea of a physical
origin of these constraints by comparing with
the study of Karliner et al. \cite{KARL1},
on the wavefunction of the ordinary string. At
distances below a fundamental length
$T^{-1/2}$, fluctuations completely dominate the
wavefunction, and it makes no sense to specify
a particular string configuration. In the
tensionless limit this behaviour should extend to all of space-time.

Another sign of a drastic reduction
of the spectrum is the scaling argument of Atick and
Witten \cite{ATIC1}, which indicates
that the short-distance degrees of freedom of string
theory are much fewer than in particle theory. In our framework this would mean
additional structure in multi-string Hilbert
spaces. Our result is then also consistent
with the selection rules found by
Gross \cite{GROS1}. In the $T \to 0$ limit amplitudes are
given by polarizations (spin) in the
scattering plane, i.e. the plane defined by the
relative momenta. Other polarization
directions do not affect the amplitudes. The
constraints (\ref{phys}) imply that no
spin is allowed for a single tensionless string, but
on the other hand there is no relative
momentum in this single string Hilbert space.

We have found that our canonical light-cone
quantization only preserves conformal symmetry
if the spectrum is severely restricted to
states invariant under general space-time
coordinate transformations. At present
we lack an explicit construction of a non-trivial
spectrum, but we believe that this result
connects well with ideas on unbroken general
covariance, and few short-distance degrees of freedom in string theory.
\bigskip
\bigskip

\bfl
{\bf Acknowledgement:} We gratefully acknowledge
numerous discussions with Georgios
Theodoridis and Hans Hansson.
\efl

\eject

\end{document}